# Analysis of TL and OSL kinetics of lithium aluminate

A. Twardak[1], P. Bilski[1], B. Marczewska[1], W. Gieszczyk[1]

[1]Institute of Nuclear Physics (IFJ), Polish Academy of Science, 31-342 Krakow, Poland

**Abstract**

Lithium aluminate ($LiAlO_2$) polycrystalline material showing high OSL sensitivity and linear dose response was prepared at IFJ Krakow. The kinetic parameters of OSL and TL processes were evaluated using various experimental techniques: LM-OSL deconvolution, TL glow-curve deconvolution, variable heating rate and isothermal decay. The OSL signal was found to consist of four components, one of them exhibiting a very slow decay. The TL glow-curve possesses two apparent peaks at approximately 85 $^oC$ and at 165 $^oC$, which both seem to follow first order kinetics. These peaks seem to have a composite structure and as many as six peaks were found in the glow-curve.

**Introduction**

Lithium aluminate ($LiAlO_2$) is one of the compounds considered as a potential materials for application in optically stimulated luminescence (OSL) dosimetry. $LiAlO_2$ contains natural lithium, consisting partly of Li-6 isotope, so that it can find application in measurement in neutron fields. Li-6 has a very high cross section for reaction with thermal neutrons: $^6Li + n -> {}^4He + {}^3H$. This is one of advantages of this material over the aluminum oxide doped with carbon ($Al_2O_3$:C), which is commonly used in OSL dosimetric systems (Landauer). Additionally, the effective atomic number of $LiAlO_2$ (10.7) is somewhat lower than $Al_2O_3$ (11.3), what results in better tissue equivalence. Lithium aluminate was for the first time studied with respect to OSL properties by Mittani et al. (2008). The prepared terbium doped $LiAlO_2$ was reported to show OSL sensitivity more than ten times lower than $Al_2O_3$:C and linear dose response characteristic from 2 mGy to 5 Gy. OSL of differently doped $LiAlO_2$ was investigated also by Dhabekar et al. (2008a), but they reported sensitivity 200 times lower than that of $Al_2O_3$:C.

The samples studied in the mentioned publications were prepared with methods of sintering and combustion synthesis, so without full melting of the material and single crystal growth. A big improvement was achieved recently by the group from KAERI (Daejon, Korea), who attempted to grow lithium aluminate single crystals using the Czochralski method and found that undoped lithium aluminate exhibits high OSL sensitivity, significantly exceeding that of $Al_2O_3$:C (Lee et al, 2012). The dose response characteristic was found to be linear in the range from 90 mGy to 2.25 Gy. The main drawback is high fading: 17 % over 10h .

Some studies of thermoluminescence properties of lithium aluminate were also reported. Dhabekar et al. (2008b) presented TL glow-curves of $LiAlO_2$: Ce and $LiAlO_2$: Mn. Manganese doped lithium aluminate TL properties were also demonstrated by Teng et al. (2010). Thermoluminescent glow-curve of undoped $LiAlO_2$ was presented by Lee et al. (2012). Recently Sadel et al. (2013) has reported $LiAlO_2$ efficiency for protons.

The papers by Lee at al. (2012, 2013) are focused mainly on the general characterization of luminescence of lithium aluminate and on its dosimetric properties. There was so far no attempt to analyze the kinetic parameters of this newly developed material. The goal of the present work was therefore to study kinetic parameters of both TL and OSL processes of $LiAlO_2$ using various experimental methods. The investigations were realized with undoped $LiAlO_2$ samples manufactured at the IFJ Kraków, which main features are very similar to those reported by Lee at al. (2012).

**Materials and methods**

Lithium aluminate samples were prepared at the IFJ PAN Kraków. The commercially available high-purity powder of lithium aluminate was melted in an iridium crucible in the Micro-Pulling-Down crystal growth system ($\mu$-PD) from Cyberstar (France) in the Ar atmosphere. The $\mu$-PD furnace is equipped with the radio frequency heating system enabling reaching temperatures up to 2100°C. The obtained polycrystalline melts were crushed into powder form. Samples with approximately 15 mg mass were further used in all measurements. Before measurements all samples were annealed at 450°C for 240 s in the reader.

Thermoluminescence glow-curves were registered using Risø DA-20 TL/OSL reader (Risø DTU, Denmark). DA-20 reader detection system is equipped with band pass filter U-340 (Hoya). It allows registering light with the wavelengths from 300 to 400 nm (UV range). This range fits exactly the OSL emission spectra of $LiAlO_2$ with peaks at 332 nm, 344 nm and 361 nm (Lee et al. 2013). Before each measurement samples were annealed in the reader at 450 °C for 240 s. This procedure ensures erasing any possible residual signal. The irradiations were conducted using beta emitter (Sr-90/Y-90) built in DA-20 reader. Standard heating rate was 5 °C/s. For optical stimulation 28 blue LEDs (peak emission at 470 nm) were used. Continues wave OSL (CW-OSL) decay curves were registered using 90% of total LEDs power, which is 80 mWcm$^{-2}$ at a sample surface. Linearly modulated optically stimulated luminescence (LM-OSL) curves were measured using linear ramping (Bulur, 1996) from 0 to 100%.

**Results and discussion**

The obtained $LiAlO_2$ samples show very high OSL sensitivity. The CW-OSL intensity per unit mass of $LiAlO_2$ was about 10 times higher than that of $Al_2O_3$:C. The measured dose-response was found to be linear in the tested dose range from 0.5 mGy to 0.5 Gy. These results, which are in agreement with the findings of Lee et al. (2012, 2013), seem to indicate that the samples studied in this work have similar characteristics as the Korean material. The obtained results should be therefore also relevant to the Korean samples.

The LM-OSL decay curves were registered during 400 s after irradiation with the dose of 700 mGy. Background of a non–irradiated sample was measured in the same conditions and subtracted from the collected data. LM-OSL curve was fitted with linear combination of the first order kinetics equation (Kitis and Pagonis 2008), using spreadsheet software package Microsoft Excel with add-in utility – Solver (Afouxenidis et al. 2012). To characterize the LM-OSL, the peak shape equation given by Kitis and Pagonis (2008) was used:

$$I(t) = 1.6487\, I_m \frac{t}{t_m} \exp\left(\frac{-t^2}{2t_m^2}\right) \quad (1)$$

where $I(t)$ is a luminescence intensity as a function of time ($t$), $I_m$ is maximum peak intensity. Time corresponding to this peak maximum ($t_m$) depends on detrapping probability ($b$) and total illumination time ($T$):

$$t_m = \sqrt{\frac{T}{b}} \quad (2)$$

Choi et al. (2006) presented a method of calculation of photoionization cross section ($\sigma$). Previously estimated detrapping probability is proportional to photoionization cross section maximum stimulation photon flux ($J$):

$$b = J\sigma \quad (3)$$

For parameters used in experiment, stimulation wavelength of 470 nm and power density of 72 mWcm$^{-2}$, the calculated photon flux was $1.7 \cdot 10^{17}$ s$^{-1}$cm$^{-2}$. Curve fitting studies have shown that four LM-OSL components are a minimum to achieve an acceptable fit (Figure of Merit - FOM% ~ 6%): three to fit the main part of LM-OSL curve (C1 – C3) and the last one for the slowly decreasing tail (Fig. 1). For each component the time corresponding to peak maximum was established. It allows calculating photoionization cross section ($\sigma$), which is related to detrapping probability, but independent from measurement conditions (Choi et al. 2006). The obtained results are presented in Table 1. Photoionization cross section is the highest for the first component and decreases by the order of magnitude for the subsequent one. The first component has also the smallest contribution to the whole LM-OSL signal (17.7%), while light outputs of C2-C4 are comparable (26.0 – 28.8 %).

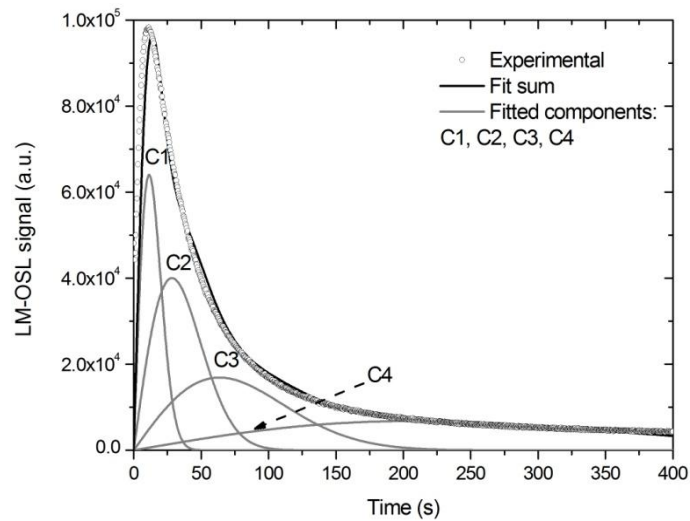

Fig. 1. Lithium aluminate LM-OSL signal decomposed into four components.

Table 1. The parameters estimated from LM-OSL curves: time of maximum ($t_{max}$), photoionization cross sections, contribution to the total signal.

| Component | $t_{max}$ (s) | σ (cm$^2$) | Contribution (%) |
|---|---|---|---|
| Fast (C1) | 11.4 | $1.6 \cdot 10^{-17}$ | 17.7 |
| Medium (C2) | 28.3 | $2.6 \cdot 10^{-18}$ | 27.5 |
| Slow (C3) | 67.5 | $4.6 \cdot 10^{-19}$ | 26.0 |
| Very slow (C4) | 206.8 | $4.9 \cdot 10^{-20}$ | 28.8 |

Thermoluminescent glow-curves were registered in the range 30 °C – 450 °C. Lithium aluminate possesses at least two easily distinguishable peaks: one with maximum at approximately 85 °C and the second at 165 °C. Temperatures of maximum TL intensity of these peaks do not depend on dose, what suggests first order kinetics (Pagonis et al. 2006).

The first method used to establish trap parameters corresponding to each peak was the variable heating rate method (VHR). This method is based on the effect, that with increase of the heating rate, the temperature of a TL maximum shifts towards higher temperatures. The simple transformation of the first order kinetic equation allows obtaining a linear relation (Pagonis et al. 2006):

$$\ln\left(\frac{T_m^2}{\beta}\right) = \frac{E}{kT_m} + \ln\left(\frac{E}{sk}\right) \quad (4)$$

where $T_m$ is temperature corresponding with peak maximum and $k$ is Boltzman constant. Plot of $\ln\left(\frac{T_m^2}{\beta}\right)$ as a function of $\frac{1}{kT_m}$ can be fitted with linear function. The slope of this function represents trap depth ($E$) and from the intersection frequency factor ($s$) can be calculated (see Fig. 2). The measurements were realized with 6 various heating rates (β): 1, 2, 4, 6, 8, 10 °C/s, after exposure to a dose of 350 mGy. The data obtained from measurements were corrected for the temperature lag using procedure described in details by Kitis and Tuyn (1998). The calculated trap parameters are presented in the Table 2 (columns 1 and 2).

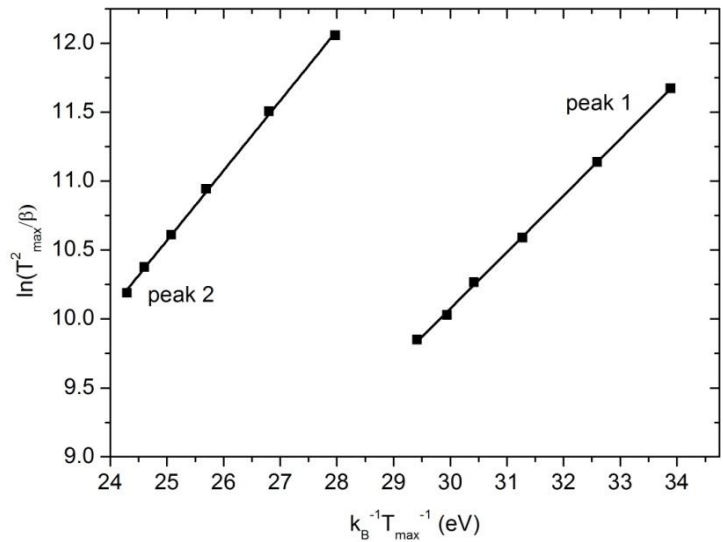

Fig. 2. Plot of $\ln(T^2_{max}/\beta)$ versus $1/(k_B T_{max})$ obtained using various heating rates for peak 1 and peak 2 of lithium aluminate.

The second method applied to establish trap parameters was deconvolution of TL glow-curves. The curves were registered with a heating rate of 5 °C/s after irradiation with a dose of 350 mGy. The obtained glow-curves were deconvoluted into single first order kinetic peaks using program GlowFit (Puchalska and Bilski, 2006). The deconvolution of a glow-curve with unknown *a priori* number of peaks, carries always a risk of over-parameterization of a model, as increased number of peaks obviously entails a numerically better fit, which not necessarily has a real physical meaning. In the present case, at first we established the minimum number of peaks needed to obtain an acceptable fit (parameterized by the FOM). This number was found to be three, with peak 1 consisting of two strongly overlapped peaks (FOM=4.4%) (see Fig. 3a). However, the shape of peak 2 suggests that it also may be a superposition of two peaks and additionally two low-intensity peaks may be observed for high-temperatures (Fig. 3b). Assuming presence of six peaks, a much better fit (FOM 2.9%) was achieved. The obtained peak parameters in both cases are presented in the Figure 3. Comparing values of activation energy with those calculated with the VHR method, one can see a big discrepancy. The reason is that the VHR method is unable to resolve the composite peak 1 into components. When we forced the GlowFit program to fit peak 1 with only one component (FOM ~ 10%), an acceptable agreement in values of *E* (and for peak 2 also for frequency factor *s*) was achieved (Table 2, columns 3, 4 and 5).

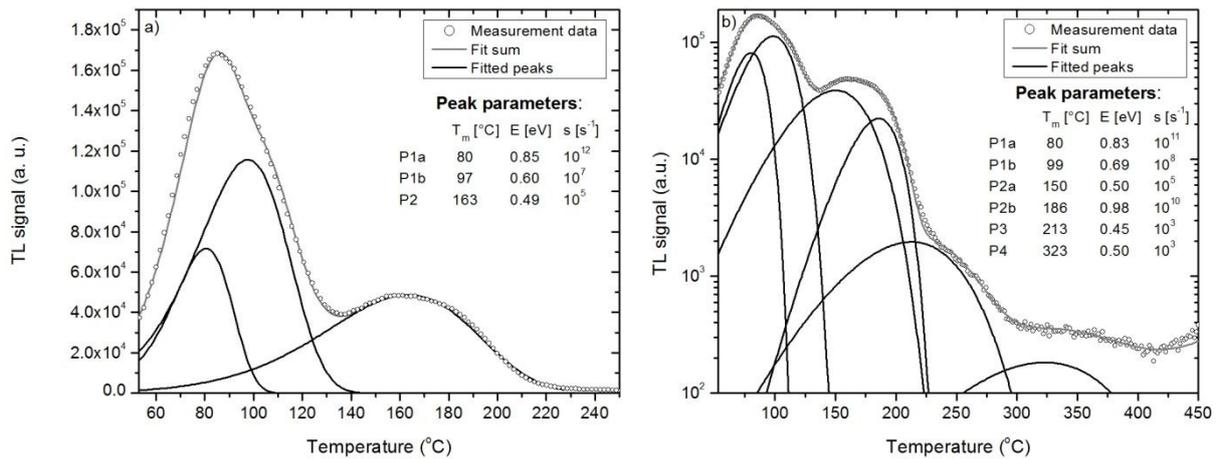

Fig. 3. Deconvolution of lithium aluminate glow curve: a) with three peaks (FOM% = 4.4%), b) with six peaks (FOM=2.9%).

The third method applied to determine trap parameters was the isothermal decay readout (ID) (Pagonis et al. 2006). For the first order kinetics, the isothermal decay curves are exponential functions of time (*t*):

$$I = I_0 \exp(-t/\tau) \qquad (5)$$

where $\tau$ is a peak lifetime, which depends on temperature ($T_i$) of ID readout:

$$\tau = s^{-1} \cdot exp\left(\frac{E}{kT_i}\right) \qquad (6)$$

The logarithm of *I* is therefore a linear function of time with the slope (1/$\tau$). Taking the natural logarithm of the parameter $\tau$ obtained for different readout temperatures $T_i$ and plotting it versus $\frac{1}{kT_i}$ should produce a straight line with the slope *E*:

$$\ln(\tau) = \ln(s^{-1}) + \frac{E}{kT_i}$$

The samples were irradiated with the dose of 350 mGy. TL glow-curves were registered using slow heating rate of 1 °C/s in order to determine precisely temperature of maximum TL signals for both peaks. The obtained temperatures were 68 °C and 138 °C for peak 1 and peak 2, respectively. ID measurements were conducted at the decay temperatures ($T$): 45, 50, 55 and 60 °C (peak 1) and 125, 130 and 135 °C (peak 2). The ID signals were registered for 600 s. Heating rate up to readout temperature was 1 °C/s, to make sure that the whole sample is heated uniformly. Before ID measurement, the samples were preheated to 40 °C (peak 1) and 120 °C (peak 2) to erase low temperature components of TL glow-curve. It was especially important for investigation of the second peak to remove the contribution of the first peak. Using the collected data, graphs of natural logarithm of ID intensity versus time were plotted. Figure 4 presents an example of the received isothermal decay curve measured at 50 °C. For the first peak several initial data points, where curve was still rising (first 10 s of readout), were skipped. It is apparent that the data are not linear with time. This means that either the studied process is not first order or the contributions from other peaks are significant. As other measurements indicate that TL processes undergo first order kinetics, we assumed the latter explanation. To fit the peak 1 decay curve, three linear functions were needed (Fig. 4). We suppose that only the first, fast component represent genuine peak 1, while two others long-live components are contributions from peaks at higher temperatures, which can be seen in Fig. 3b. Similarly, for peak 2, two components were found, with the fast one apparently being a contribution from the more intensive peak 1. Figure 5 presents plots of $\ln(\tau)$ versus $1/kT$ for both peaks. The obtained values of parameters are listed in the Table 2 (columns 6 and 7). The activation energy values approximately agree with those produced by the VHR method and by deconvolution with the simplified two-peak model. The value of the frequency factor for peak 2 significantly differs from those obtained with other methods. The ID method seems to be the least reliable due to difficult to avoid influence of satellite peaks.

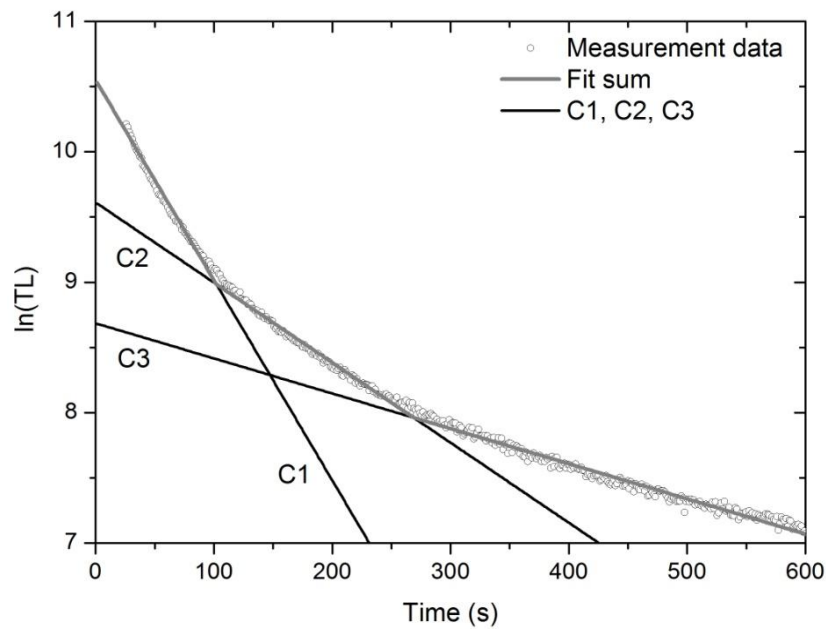

Fig. 4. Isothermal decay curve of peak 1 measured at 50 °C fitted with three components.

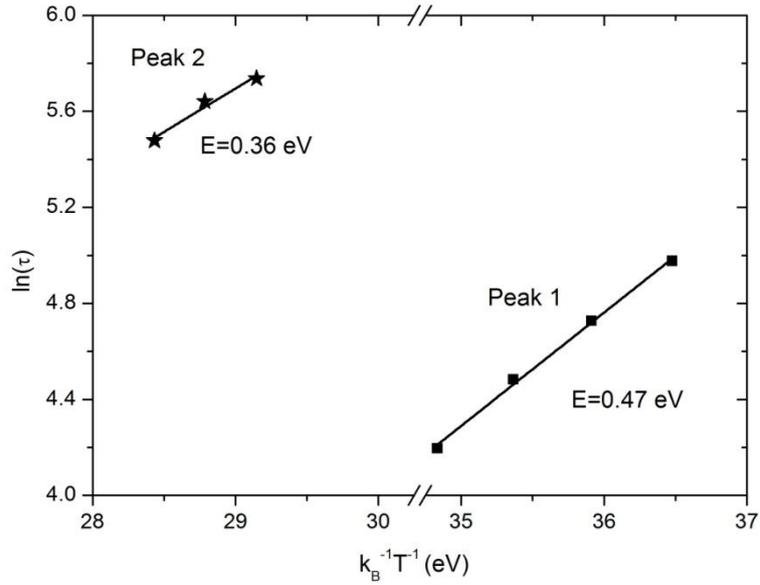

Fig. 5. Plot of natural logarithm of peak's lifetime ( ln(τ) ) versus 1/kT for peak 1 and 2.0. Energy values calculated from linear fits.

Table 2. Parameters of lithium aluminate traps calculated using, various heating rates method, isothermal decay and deconvolution ( two peaks model).

|  | Various heating rates | | Deconvolution | | | Isothermal decay | |
|---|---|---|---|---|---|---|---|
|  | E (eV) | s (s$^{-1}$) | T$_{max}$ ($^o$C) | E (eV) | s (s$^{-1}$) | E (eV) | s (s$^{-1}$) |
| **Peak 1** | 0.41 | ~ 10$^4$ | 89 | 0.52 | ~ 10$^6$ | 0.47 | ~ 10$^4$ |
| **Peak 2** | 0.51 | ~ 10$^5$ | 164 | 0.50 | ~ 10$^5$ | 0.36 | ~ 10$^2$ |

The gathered experimental data and the performed kinetic analysis does not allow to draw conclusions upon the nature of the trapping sites responsible for TL/OSL processes. Nevertheless, as the observed luminescent effects occur in the undoped LiAlO$_2$, we suppose that they are caused by structural defects and probable candidates for trapping sites seem to be oxygen vacancies: like F$^+$ centers (oxygen vacancy trapping an electron) and F$^0$ centers (oxygen vacancy trapping two electrons).

**Conclusions**

Polycrystalline samples of undoped lithium aluminate showing high OSL sensitivity and linear dose response were obtained and investigated. The kinetic parameters of OSL and TL processes were evaluated using various experimental techniques. OSL signal was found to consist of four components, one of them exhibits a very slow decay. The TL glow-curve possesses two apparent peaks, which seem to follow first order kinetics. The isothermal decay and the variable heating rate method gave for both peaks activation energy values around 0.4 - 0.5 eV. Similar values were obtained also with the deconvolution, if the glow-curve was fitted with only two peaks. The real number of TL peaks is however higher: peak 1 has certainly a composite structure and probably the same regards also to peak 2. The isothermal decay and the variable heating rate methods were unable to resolve so strongly overlapped components. The performed analysis are just the first step in investigating of trapping process in lithium aluminate and further work is needed for better understanding of TL/OSL mechanism in this material.

**Acknowledgements**

This work was supported by the National Science Centre (No. DEC-2012/05/B/ST5/00720). A. Twardak has been partly supported by the EU Human Capital Operation Program, Polish Project No. POKL.04.0101-00-434/08-00.